\documentclass[journal=jacsat, manuscript=article]{achemso}

\usepackage{graphicx}
\usepackage[T1]{fontenc}
\usepackage[utf8]{inputenc}
\usepackage{mathptmx}
\usepackage{etoolbox}

\usepackage{siunitx}
\usepackage{mhchem} 
\usepackage{float}
\usepackage{hyperref}
\usepackage{placeins}

\newcommand\sref[2]{\hyperref[#1]{\ref*{#1}(#2)}}
\newcommand\sample{FA$_{0.9}$Cs$_{0.1}$PbI$_{2.8}$Br$_{0.2}$}



\title{Spin-dependent exciton-exciton interactions in a mixed lead halide perovskite crystal}
\date{\today}

\author{Stefan~Grisard}
\email{stefan.grisard@tu-dortmund.de}
\affiliation{Experimentelle Physik 2, Technische Universit\"at Dortmund, 44221 Dortmund, Germany}

\author{Artur~V.~Trifonov}
\affiliation{Experimentelle Physik 2, Technische Universit\"at Dortmund, 44221 Dortmund, Germany}

\author{Thilo~Hahn}
\affiliation{Institute of Solid State Theory, University of Münster, 48149 Münster, Germany}

\author{Tilmann~Kuhn}
\affiliation{Institute of Solid State Theory, University of Münster, 48149 Münster, Germany}

\author{Oleh Hordiichuk}
\affiliation{Laboratory of Inorganic Chemistry, Department of Chemistry and
Applied Biosciences, ETH Z\"urich, CH-8093 Z\"urich, Switzerland}
\alsoaffiliation{Laboratory for Thin Films and Photovoltaics, Empa-Swiss Federal Laboratories
for Materials Science and Technology, CH-8600 D\"ubendorf, Switzerland}

\author{Maksym~V.~Kovalenko}
\affiliation{Laboratory of Inorganic Chemistry, Department of Chemistry and
Applied Biosciences, ETH Z\"urich, CH-8093 Z\"urich, Switzerland}
\alsoaffiliation{Laboratory for Thin Films and Photovoltaics, Empa-Swiss Federal Laboratories
for Materials Science and Technology, CH-8600 D\"ubendorf, Switzerland}

\author{Dmitri~R.~Yakovlev}
\affiliation{Experimentelle Physik 2, Technische Universit\"at Dortmund, 44221 Dortmund, Germany}

\author{Manfred~Bayer}
\affiliation{Experimentelle Physik 2, Technische Universit\"at Dortmund, 44221 Dortmund, Germany}

\author{Ilya~A.~Akimov}
\email{ilja.akimov@tu-dortmund.de}
\affiliation{Experimentelle Physik 2, Technische Universit\"at Dortmund, 44221 Dortmund, Germany}

\begin{document}    


\begin{abstract}
We investigate the two-pulse photon echo response of excitons in the mixed lead halide perovskite crystal \sample in dependence on the excitation intensity and polarization of the incident laser pulses. 
Using spectrally narrow picosecond laser pulses, we address localized excitons with long coherence times $T_2 \approx \qty{100}{\pico\second}$. 
This approach offers high sensitivity for the observation of excitation-induced changes in the homogeneous linewidth $\Gamma_2=2\hbar/T_2$ on the $\mu$eV scale. 
Through intensity-dependent measurements, we evaluate the increase of $\Gamma_2$ by 10~$\mu$eV at an exciton density of 10$^{17}$~cm$^{-3}$ being comparable with the intrinsic linewidth of \qty{14}{\mu\eV}.
We observe that the decay of the photon echo and its power dependence are sensitive to the polarization configuration of the excitation pulses, 
which indicates that spin-dependent exciton-exciton interactions contribute to excitation-induced dephasing. 
In cross-linear polarization, the decay is faster and its dependence on exciton density is stronger as compared to the co-polarized configuration. 
Using a two-exciton model accounting for different spin configurations we are able to reproduce the experimental results.

\end{abstract}    

\maketitle

\section{Introduction}
Hybrid organic-inorganic perovskites such as \ce{FA_{1-x}Cs_{x}PbI_{3-y}Br_y} attract profound attention due to their potential for stable and efficient photovoltaic or light-emitting devices with adjustable 
properties~\cite{min_efficient_2019, chen_advances_2021, zheng_development_2022, docampo_long_term_2016}.
Moreover, perovskite semiconductors offer promising properties for spintronic applications~\cite{wang_spin_optoelectronic_2019, kim_chiral_induced_2021}, where optical methods can be used to establish long-lived spin polarizations~\cite{kirstein_lande_2022}.
Understanding the dynamics of optically excited carriers and the role of their spin-dependent interactions is therefore of high relevance for the optimization of spin transport and light emission properties.

The Coulomb interaction between photo-excited electron-hole pairs leads to the formation of excitons that dominate the optical response near the band edge at low temperatures. 
The transient properties of the exciton resonance upon photo-excitation represent a sensitive probe for Coulomb interactions in semiconductors. 
In this context, extensive studies were performed in conventional semiconductors such as GaAs and quantum well heterostructures using transient pump-probe~\cite{trifonov_nanosecond_2019, trifonov_exciton_2020} and four-wave mixing (FWM) spectroscopy~\cite{shah_2010}, where the important role of exciton-exciton interactions 
is demonstrated~\cite{ciuti_role_1998, gribakin_exciton-exciton_2021}. 
The broadening of the homogeneous linewidth $\Gamma_2$ with the increase of exciton density $n_X$ was attributed to excitation-induced dephasing (EID). 
Broadenings by about 100~$\mu$eV were observed at $n_X\approx 10^{15}$~cm$^{-3}$ in bulk GaAs~\cite{honold_collision_1989} and $10^{16}$~cm$^{-3}$ in ZnSe~\cite{wagner_coherent_1997}.  
Numerous FWM studies conducted on GaAs-based structures have shown that polarization-sensitive measurements provide rich information on exciton interactions involving the optical excitation of bound biexcitons~\cite{yaffe_polarization_1993, haase_coherent_1998, bristow_polarization_2009} as well as higher energy unbound biexciton states~\cite{mayer_evidence_1994, singh_polarization-dependent_2016, svirko, turner_coherent_2010, karaiskaj_two-quantum_2010}. 
For disordered systems where excitons are localized at potential fluctuations it was shown that the coherence time $T_2\propto\Gamma_2^{-1}$ increases~\cite{steel_photon_2002} while the inhomogeneous broadening of the binding energy and the continuum edge of biexciton states manifests itself as EID~\cite{langbein_coherent_1998, langbein_biexcitonic_2002}.

Recent studies in two-dimensional WSe$_2$ and MoSe$_2$ reported an increase of $\Gamma_2$ by about \qty{1}{\milli\eV} for exciton densities of 10$^{12}$cm$^{-2}$, which is larger as compared to conventional semiconductor quantum well structures when normalized to the exciton Bohr radius~\cite{moody_intrinsic_2015, boule_coherent_2020}.
Theoretical approaches using local fields to account for exciton-exciton interactions in a MoSe$_2$ monolayer were used to explain the observation of a destructive photon echo in the six-wave mixing signal~\cite{hahn_destructive_2022}.

In bulk perovskite semiconductors, the exciton binding energy and Bohr radius are comparable to those in conventional GaAs and ZnSe semiconductors. Therefore, similar strengths of exciton-exciton interactions are expected~\cite{galkowski_determination_2016}. 
However, the magnitude of interactions and in particular their spin dependences are controversial. 
Polarization-resolved pump-probe studies on \ce{CsPbBr3} claimed a pronounced excitation-induced circular dichroism due to spin-dependent Coulomb interactions with a magnitude of about \qty{12}{\milli\eV} at moderate exciton densities of roughly \qty{2e17}{\centi\meter^{-3}}~\cite{zhao_transient_2020},
which is significantly larger than the excitation-induced shifts reported in Refs.~\citenum{march_four-wave_2017, masharin_room-temperature_2023}. 
It has been demonstrated that at low temperatures the inhomogeneous broadening of optical transitions dominates over $\Gamma_2$ in organic-inorganic perovskites~\cite{grisard_long-lived_2023, nazarov_photon_2022, trifonov_photon_2022}. 
In this case, the nonlinear optical response is given by photon echoes which can be used to extract rich information on the homogenous linewidth of the exciton and the role of EID.

In this work, we study the polarization and intensity dependence of the nonlinear optical response of excitons in a \sample single crystal at $T = \qty{2}{\kelvin}$ using time-resolved photon echo spectroscopy. 
The use of spectrally narrow picosecond pulses for excitation of localized excitons with a homogeneous linewidth of \qty{14}{\mu\eV} ensures an exceptional sensitivity for interaction effects on the $\mu$eV energy scale. 
We observe that the decay of the photon echo is shorter for linearly cross-polarized excitation pulses as compared to co-polarized pulses. 
Moreover, the broadening of the homogeneous linewidth is more pronounced in cross-polarized configuration when increasing the excitation intensity accompanied by an increase in exciton density.
To gain insight into this effect, we continuously vary the linear polarization angle between the excitation pulses, which allows us to exclude contributions of bound biexciton states with zero total spin to the polarization-sensitive exciton linewidth. 
The results are interpreted in terms of spin-sensitive EID where excited two-exciton states with parallel spins influence the measured exciton linewidth. 
The maximum observed excitation-induced broadening amounts to \qty{10}{\mu\eV} at an exciton density of \qty{1.4e17}{\centi\meter^{-3}}. 
At this energy scale, spin-dependent interactions are found to play a decisive role.

\section{Sample and Methods}
The studied sample is a solution-grown single crystal of the composition \sample. The same sample was previously studied
using transient four-wave mixing (FWM) spectroscopy in Ref.~\citenum{grisard_long-lived_2023}, where further details on 
the growth technique can be found.  \\
\sample \,exhibits a direct band gap of $\approx \qty{1.52}{\eV}$ at $T = \qty{2}{\kelvin}$ with 
a pronounced exciton feature shifted towards lower energies by the binding energy of \qty{14}{\milli\eV}~\cite{kopteva_giant_2023}. 
In Ref.~\citenum{grisard_long-lived_2023}, the exciton coherence time $T_2$ was found
to be exceptionally long in the order of \qty{100}{\pico\second}, which was
associated with the localization of excitons on spatial modulations of the bandgap on the nanometer scale. 

\begin{figure*}
    \centering
    \includegraphics[scale = 1]{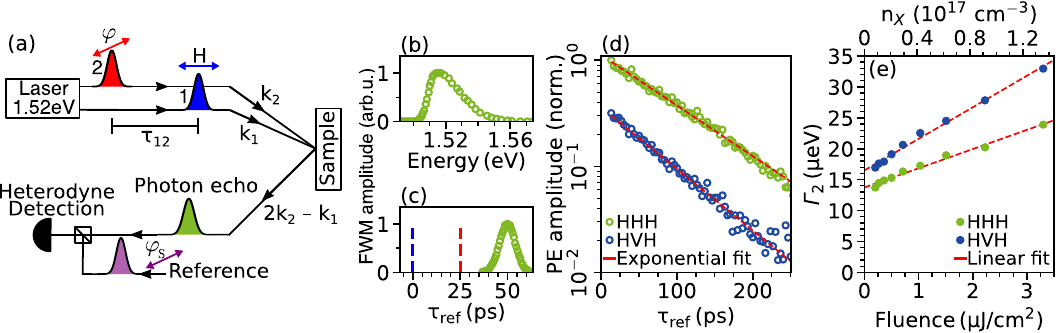}
    \caption{(a) Schematic picture of the experimental setup for time- and polarization-resolved four-wave mixing 
    spectroscopy as described in the main text. (b) Four-wave mixing spectrum of the studied \sample crystal exhibiting a broad peak at the exciton resonance. 
    The spectrum is measured for temporally overlapping first and second pulses, i.e. 
    $\tau_{12} = 0$. All subsequent measurements are performed for a central photon energy of 1.52\,eV. 
    (c) Exemplary time-resolved photon echo pulse. Vertical dashed lines show the temporal positions of 
    the first and second pulses. (d) Decay of the photon echo amplitude as a function of $\tau_\text{ref} = 2\tau_{12}$ for 
    a total laser fluence of \qty{0.3}{\mu\joule\per\centi\meter^2} measured in the polarization configurations HHH and HVH. 
    Red lines are fits to single exponentials. (e) Homogenous linewidth $\Gamma_2$ measured in HHH and HVH as a function of the total laser fluence. 
    Red dashed lines are fits to linear functions.}
    \label{fig: fig01}
\end{figure*}

We use a polarization- and time-resolved FWM technique that is schematically illustrated 
in Fig.~\sref{fig: fig01}{a}. A modelocked Ti:Sa laser is used as a source of pulse trains with a repetition 
rate of \qty{75.75}{\mega\hertz}
and pulse duration of \qty{3.3}{\pico\second} (associated with the intensity profile; spectral width  $\approx\qty{0.4}{\milli\eV}$). 
The pulses are split into two excitation paths with an adjustable 
delay of $\tau_{12}$ and respective 
wavevectors $\mathbf{k}_1$ and $\mathbf{k}_2$. 
The sample is placed in a helium bath cryostat where it is cooled down to the temperature of \qty{2}{\kelvin}.
The pulses are focussed to a spot size of roughly \qty{150}{\mu\meter} on the sample surface. 
We study the photon echo response resulting from the third-order polarization 
$P^{(3)}\propto E_1^*E_2^2$ emitted in the phase-matched direction $2\mathbf{k}_2 - \mathbf{k}_1$, 
where $E_i$ are the electric field amplitudes of the optical pulses and $\mathbf{k}_i$ their respective wavevectors. 
The central photon energy of the picosecond laser pulses is tuned to the exciton 
resonance at $\approx\qty{1.51}{\eV}$, which dominates the 
FWM spectrum of the system as shown in Figure~\sref{fig: fig01}{b}.
We stress that the FWM spectrum is strongly inhomogeneously broadened where the inhomogenous broadening ($\approx\qty{16}{\milli\eV}$) exceeds the homogeneous exciton linewidth \qty{14}{\mu\eV} by three orders of magnitude. We thus deal with a system of localized, and thus energetically discrete, exciton states in a disordered environment. 

The magnitude of the FWM electric field is detected using the time-resolved heterodyne technique, where the 
signal pulse is brought to interference with a strong reference pulse with adjustable delay $\tau_\text{ref}$ relative to the first excitation pulse. Figure~\sref{fig: fig01}{c} presents an exemplary time-resolved photon echo response of the system where 
the delay of first and second 
pulse is set to \qty{25}{\pico\second} inducing a photon echo pulse centered at \qty{50}{\pico\second}. 
A measurement of the decay of the photon echo amplitude $\exp(-2\tau_{12}/T_2)$ 
as a function of $\tau_{12}$ allows us to measure the 
decoherence time $T_2$ corresponding to the homogeneous linewidth $\Gamma_2 = 2\hbar / T_2$. 
The relative polarization of optical pulses and the detected polarization as well as the 
excitation fluence are controlled by combinations of wave plates and Glan-Taylor polarizers. 

\section{Excitation-induced dephasing}
We aim to quantify the contribution of EID to the homogenous linewidth of excitons in \sample. 
For this purpose, we measure the exciton coherence time $T_2$ as a function of the applied laser fluence.
The fluence of the first pulse is constant at \qty{0.1}{\mu\joule\per\centi\meter^2}, while the fluence of the second pulse is varied. 
First, all pulses are horizontally polarized and the signal is detected in the same polarization (configuration   
denoted as HHH in the following).  
Figure~\sref{fig: fig01}{d} shows the decay   
of the photon echo amplitude as a function of $\tau_\text{ref} = 2\tau_{12}$ for a total laser fluence of  
\qty{0.3}{\mu\joule\per\centi\meter^2}. A fit to an exponential function 
yields the decoherence time of $T_2 = \qty{91(1)}{\pico\second}$ corresponding to a narrow 
homogeneous linewidth of $\Gamma_2 = 2\hbar / T_2 = \qty{14}{\mu\eV}$,  
thus reproducing the results presented in Ref.~\citenum{grisard_long-lived_2023}.   
The homogeneous linewidth increases linearly with the applied laser fluence with a 
slope of \qty{3}{\mu\eV \per \mu\joule \centi\meter^{-2}} as 
shown in Figure~\sref{fig: fig01}{e}. 
On top of Figure~\sref{fig: fig01}{e}, we recalculated the total laser fluence to the exciton density $n_X$, for which 
we assumed an absorption coefficient of $10^{4}$\,cm$^{-1}$~\cite{wright_intrinsic_2020}.
The maximum broadening at the highest estimated exciton density of \qty{1.4e17}{\centi\meter^{-3}} amounts to approximately $10\,\mu$eV.

Surprisingly, a pronounced horizontally polarized photon echo signal is observed when the first and second pulses 
are linearly cross-polarized (configuration denoted as HVH), which is not expected from non-interacting 
excitons~\cite{yaffe_polarization_1993, poltavtsev_polarimetry_2019}.  
The decay of this signal is with $T_2 = \qty{75(1)}{\pico\second}$ notably shorter than observed in 
HHH, Figure~\sref{fig: fig01}{b}.
This difference in decoherence times increases when increasing the total laser fluence as we 
present in Figure~\sref{fig: fig01}{e}. Here, the increase of the linewidth in HVH exhibits a larger slope of 
\qty{5}{\mu\eV \per \mu\joule \centi\meter^{-2}}. In this way, the ratio between the decoherence rates in HVH and HHH increases from $\approx 1.2$ for zero fluence (linearly extrapolated) to $\approx 1.4$ in the observed range of laser fluences.
Thus, both the intrinsic linewidth as well as the strength of the observed EID effect are sensitive to the 
relative polarization of the excitation pulses. This observation hints at the importance of 
the spin degree of freedom in the interaction of photo-excited excitons. 
In general, the increase of the linewidth may be related to the heating of the crystal lattice by laser excitation. 
However, a temperature-induced broadening should be independent of the polarization configuration.
To gain additional insight into the polarization dependence of the exciton resonance, we make use of the 
photon echo polarimetry technique in the following.

\section{Photon echo polarimetry}\label{sec: polarimetry}
\begin{figure*}
    \centering
    \includegraphics[scale = 1]{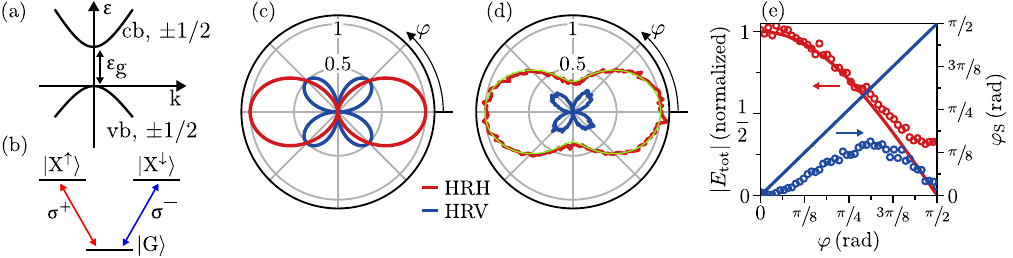}
    \caption{(a) Energy band diagram $\varepsilon(k)$ in the vicinity of the R-point of the Brillouin zone of the studied \sample\, perovskite 
    crystal. (b) V-scheme describing the polarization selection rules for optical excitation 
    of the two bright exciton states with spin $\pm1$ with circularly polarized light $\sigma^\pm$. (c) Theoretical polar dependences of the photon echo amplitude 
    in the configurations HRH and HRV (as defined in the text) assuming non-interacting excitons with a level scheme as shown in (b). (d) Experimentally 
    observed polar dependences for a delay of $\tau_{12} = \qty{20}{\pico\second}$. The green line shows a fit of the HRH dependence to a function of the form 
    $\cos^2\varphi + c$. (e) Polar dependences of the total signal amplitude $|E_\mathrm{tot}|$ 
    and the linear polarization angle $\varphi_S$ of the photon echo signal (blue). 
    Solid lines correspond to the theoretical dependences within the model of non-interacting excitons.}
    \label{fig: fig02}
\end{figure*}    

We apply the photon echo polarimetry technique as introduced in Ref.~\citenum{poltavtsev_polarimetry_2019}. 
Here, we detect the horizontally (H) and vertically (V) polarized components of the photon echo amplitude 
as a function of the relative polarization angle $\varphi$ between the horizontally polarized
first pulse and second pulse. The choice of horizontal polarization for the first
pulse is arbitrarily chosen since only the relative polarization angles are relevant.
We denote the two measurement routines as HRH and HRV in the following. 
Lead halide perovskites exhibit a band structure close to the R-point of the Brillouin zone that is shown in Fig.~\sref{fig: fig02}{a}. 
The lowest energy transition takes place between the s-type valence band formed from hybridized Pb $6s$ and halide $5p$ orbitals and the split-off band of the $p$-type conduction band resulting from unoccupied Pb $6p$ orbitals~\cite{frost_atomistic_2014, boyer-richard_symmetry-based_2016}. 
Both bands are two-fold degenerate with the spin configurations $\pm 1/2$ for electrons and holes. This results in two 
bright exciton states 
$|\text{X}^{\pm}\rangle$ with spin $\pm 1$. As illustrated by the V-scheme in Figure~\sref{fig: fig02}{b}, 
the two exciton states can be separately excited from a common ground state $|G\rangle$ using circularly polarized light. 
The theoretical dependences of the signal field amplitude in the configurations HRH and HRV are obtained 
by solving the optical Bloch equations for the V-scheme as described in 
Ref.~\cite{poltavtsev_polarimetry_2019} and SI section S1: 
\begin{subequations}
    \label{eq: rosettes_uncoupled}
\begin{align}
    \left|E_\text{HRH}\right| &\propto \cos^2(\varphi) \label{eq: HRH}\\
    \left|E_\text{HRV}\right| &\propto \frac{1}{2}\left|\sin(2\varphi)\right|. \label{eq: HRV}
\end{align}
\end{subequations}
These dependences are visualized as polar plots in Fig.~\sref{fig: fig02}{c} in red and blue, respectively. 
Experimentally, such behavior of non-interacting excitons was observed for example in CdTe quantum 
wells~\cite{poltavtsev_polarimetry_2019} or in the perovskite single crystal 
\ce{MAPbI3}~\cite{nazarov_photon_2022}. 
Two aspects should be highlighted. First, the linear polarization angle $\varphi_S$ of the photon echo signal is directly set by the polarization angle of the second pulse 
\begin{equation}
    \varphi_S = \arctan \left( \frac{E_\text{HRV}}{E_\text{HRH}}\right) = \varphi. \label{eq: phi_S}
\end{equation}
Second, the total signal amplitude 
\begin{equation}
    \left| E_\text{tot} \right| = \left(\left|E_\text{HRH}\right|^2 + \left|E_\text{HRV}\right|^2\right)^{1 / 2}\propto \left|\cos \varphi\right| \label{eq: E_tot}
\end{equation} 
vanishes in the configuration $\varphi = \pi/2$, i.e. when the excitation pulses are cross-polarized. 
This property results from the fact that the two exciton transitions are not two independent two-level systems 
but share a common ground state $|\text{G}\rangle$. The population densities $n^i$ of the three states are assumed to fulfill 
the equation $1 = n^{+} + n^{-} + n^{G}$. 
Thus, a population of one exciton species also results in a nonlinear 
response from the second species 
as a consequence of the depletion of the ground state. 
Moreover, also the spin polarization, i.e. the superposition of the two excited states gives rise to a FWM response. 
For linearly cross-polarized excitation pulses, the multiple 
contributions to the FWM process cancel each other such that the total signal amplitude is zero. 
The observation of a signal in the configuration HVH presented above is thus a clear deviation from the assumption 
of non-interacting excitons. 

To further analyze deviations of the measured polar dependences from Equations~\eqref{eq: rosettes_uncoupled} to~\eqref{eq: E_tot}, 
we introduce a delay of $\tau_{12} = \qty{20}{\pico\second}$ 
between the first and second pulse and measure the photon echo amplitude at $\tau_\text{ref} = 2\tau_{12} = \qty{40}{\pico\second}$ in the 
configurations HRH and HRV. 
The two resulting polar dependences are plotted in Fig.~\sref{fig: fig02}{d}. 
In HRH configuration, we observe a clear deviation from the $\cos^2(\varphi)$ dependence since a pronounced 
signal is observed in the cross-polarized configuration $\varphi = \pi / 2$, which amounts 
to roughly \qty{35}{\percent} of the signal amplitude in co-polarized configuration $\varphi = 0$.
As highlighted by the green line, the dependence is excellently described by a function of the form 
$\propto \cos^2\varphi + c$, where $c$ is a horizontally polarized signal component which is 
independent of the relative polarization between the first and second pulse. 
The angular dependence HRV does not qualitatively deviate from the $\left| \sin 2\varphi \right|$ dependence predicted by 
Eq.~\eqref{eq: HRV}. However, the ratio between the maximum signal strength in HRH and HRV is larger than 2, compare Eqs.~\eqref{eq: HRH} and~\eqref{eq: HRV}.  
The measured total signal amplitude as well as the linear polarization angle $\varphi_S$ as a function of $\varphi$ are compared to 
Eqs.~\eqref{eq: phi_S} and~\eqref{eq: E_tot} in Fig.~\sref{fig: fig02}{e}. Here, the most striking differences between the experimental 
observation and the picture of non-interacting excitons are given by the finite signal amplitude $|E_\mathrm{tot}|$ for $\varphi = \pi / 2$ as well as the 
strongly modified polarization angle of the signal $\varphi_S = \varphi_S(\varphi)$. 

The appearance of a FWM signal for linearly cross-polarized excitation pulses is 
often associated with the formation of multi-excitonic complexes~\cite{yaffe_polarization_1993, haase_coherent_1998, 
bristow_polarization_2009, suzuki_coherent_2016, poltavtsev_polarimetry_2019, trifonov_photon_2022, grisard_multiple_2022}. 
In particular, the binding of excitons to localized carriers leads to the formation of trions or two excitons 
may form a biexciton state. However, the polarimetric behavior in both cases is qualitatively 
incompatible with our findings as we discuss in more 
detail in SI section~S2. 
Moreover, we note that the biexciton binding energy in a similar perovskite crystal
\ce{MAPbI3} is \qty{2.4}{\milli\eV}~\cite{trifonov_photon_2022}, which we 
regard as a lower bound for the biexciton binding energy 
in \sample since localization typically leads to an increased biexciton binding energy~\cite{cundiff_semiconductor_2008}. 
Because our laser pulses are spectrally narrow, \qty{0.4}{\milli\eV}, the formation of biexcitons is excluded. 
Instead, we associate all presented observations with an influence of spin-dependent EID on the nonlinear optical response of the 
studied \sample crystal, which we theoretically analyze in the following section.

\section{Theoretical analysis of spin-dependent exciton interactions}
In this section, we present a theoretical model that accounts for spin-dependent exciton-exciton interactions, enabling us to reproduce our main experimental observations: the difference in decoherence rates in the configurations HHH and HVH (Figure~\sref{fig: fig01}{d}), as well as the qualitative shape of the polar rosettes measured in the configurations HRH and HRV (Figure~\sref{fig: fig02}{d}).
In the model, we focus on the effect of EID, assuming that the impact of excitation-induced shifts (EIS) of the exciton resonance is of minor importance. 
This simplification is justified by considering that EIS would lead to quantum beats, which are not observed. 
Moreover, the presence of macroscopic EIS is improbable in strongly inhomogeneously broadened systems as the one under study, where energy fluctuations effectively transform EIS to EID. Here, the energy spectrum of interacting excitons becomes continuous
due to the averaging of excitation-induced shifts in disordered systems or due to a large number of interacting excitons~\cite{ciuti_role_1998, langbein_biexcitonic_2002, gribakin_exciton-exciton_2021}.
Within the model presented below, we introduce effective two-exciton states that correspond to excitons probed in a local environment that depends on the excitation level. 
In this approach, we also cover potential exciton-carrier interactions, where EID arises due to the scattering of excitons by photo-excited carriers. 
The exchange interaction responsible for the spin-dependent part of EID is similar for both exciton-exciton and exciton-carrier interaction~\cite{ciuti_role_1998, gribakin_exciton-exciton_2021}.

Before we present the primary model, we briefly comment on a well-established phenomenological approach to consider excitation-induced nonlinearities
where the optical Bloch equations are expanded by a population-dependent decoherence rate $\Gamma_2 \rightarrow \Gamma_2 + \alpha n_X$~\cite{shah_2010}. Here, the interaction constant $\alpha$ measures the increase of the linewidth with increasing exciton density $n_X$.
To allow for spin-dependent exciton interactions, one may further distinguish between the 
populations of spin-up and spin-down excitons $\Gamma_2^{\pm} = \Gamma_2 + \alpha_{\uparrow\uparrow} n^{\pm} + \alpha_{\uparrow\downarrow} n^{\mp}$, 
where $n^{\pm}$ denote the spin up/down exciton populations and $\alpha_{\uparrow\uparrow}$/$\alpha_{\uparrow\downarrow}$
are interaction constants for excitons with the same or opposite spin.   
As shown in detail in SI section~S3, this approach indeed can reproduce the qualitative shape of the polar rosettes measured in the configurations HRH and HRV 
for the case $\alpha_{\uparrow\downarrow} > \alpha_{\uparrow\uparrow}$, i.e. for a stronger broadening in the presence of excitons with opposite spin. 
However, the simple approach using population-dependent decay rates is not capable of reproducing the observed polarization-dependent decoherence rates (SI section S3).  
We thus present an alternative model adopted from References~\cite{singh_polarization-dependent_2016, svirko, turner_coherent_2010, karaiskaj_two-quantum_2010} that takes 
into account EID based on correlated two-exciton states. 
The model makes statements about a polarization-dependent homogeneous linewidth and can quantitatively 
reproduce the experimentally observed polar rosettes.
\begin{figure*}
    \centering 
    \includegraphics[scale = 1]{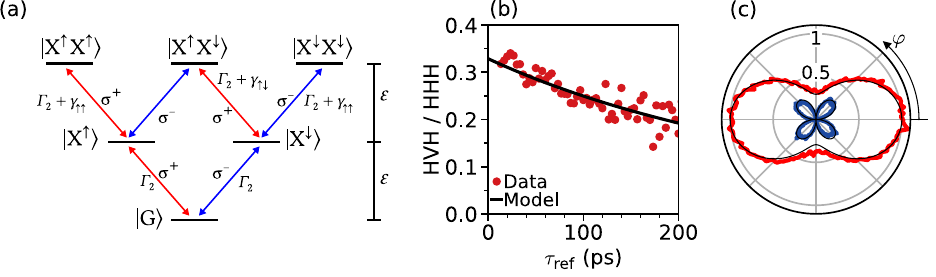}
    \caption{(a) Visualization of the two-exciton model as introduced in the text. (b) Dynamics of the ratio HVH/HHH
    as measured for a fluence of \qty{0.2}{\mu\joule\centi\meter^{-2}} and fit to the model Eq.~\eqref{eq: HHH_HVH_ratio_2X} with the estimated parameters 
    $\gamma_{\uparrow\uparrow} = \qty{5.6(5)}{\mu\eV}$ and $\nu = \num{0.30(1)}$.
    (c) Comparison of the experimental (red and blue lines) and modeled (black lines) polar rosettes, Eq.~\eqref{eq: two_X_simplifed_HRH} and Eq.\,S19, in the configurations HRH and HRV, respectively.}
    \label{fig: eid_2X_model}
\end{figure*} 

We consider a two-exciton model visualized in Figure~\sref{fig: eid_2X_model}{a} consisting of the ground state $|\text{G}\rangle$, singly excited exciton states $|\text{X}^\uparrow\rangle$ /  $|\text{X}^\downarrow\rangle$ at energy $\varepsilon$, as well as the three different spin configurations for unbound but correlated two-exciton states $|\text{X}^\uparrow \text{X}^\uparrow\rangle$, $|\text{X}^\uparrow \text{X}^\downarrow\rangle$, and $|\text{X}^\downarrow \text{X}^\downarrow\rangle$ at energy $2\varepsilon$.  
Note that higher correlated states with more than two excitons are not excited when we neglect nonlinear polarizations higher than FWM.
In such consideration, many-body effects can be introduced phenomenologically as excitation-induced shifts or modified decoherence rates of the two-exciton to one-exciton states~\cite{singh_polarization-dependent_2016}.  
As mentioned above, we solely consider modified decoherence rates $\Gamma_2 \rightarrow \Gamma_2 + \gamma$, where we further distinguish between parallel and antiparallel spins ($\gamma_{\uparrow\uparrow}$ / $\gamma_{\uparrow\downarrow}$), as indicated in Figure~\sref{fig: eid_2X_model}{a}. 
The phenomenological decoherence rates $\Gamma_2, \gamma_{\uparrow\uparrow}$, and $\gamma_{\uparrow\downarrow}$ are assumed to depend on the applied laser power. 
The power dependence of the spin-independent part of EID ($\gamma_{\uparrow\uparrow} = \gamma_{\uparrow\downarrow}$) may be associated with heating of the lattice, while the spin-dependent part ($\gamma_{\uparrow\uparrow} \neq \gamma_{\uparrow\downarrow}$) is associated with exchange interaction.
Calculating the FWM signal of the six-level system in Figure~\sref{fig: eid_2X_model}{a} leads to the interference of quantum paths involving the two-exciton to one-exciton and one-exciton to ground state transitions.  
In this way, the polarization dependence, as well as the decay dynamics of the system, is modified with respect to the one-exciton model (Figure~\sref{fig: fig02}{b}). 
Following References~\cite{singh_polarization-dependent_2016, svirko}, we construct the dipole moments of the transitions $|\text{X}^{\uparrow\downarrow}\rangle \rightarrow |\text{X}^{\uparrow\downarrow} \text{X}^{\uparrow\downarrow}\rangle$ as $\sqrt{2} (1 - \nu) \mu$ and of the transitions $|\text{X}^{\uparrow\downarrow}\rangle \rightarrow |\text{X}^\uparrow \text{X}^\downarrow\rangle$ as $\mu$, where $\mu$ is the dipole moment of the ground state to one-exciton transitions. 
The parameter $\nu$, with $0\leq \nu \leq 1$, accounts for the relative importance of phase-space filling to the FWM response.
In the case $\nu = 0$, phase-space filling is absent such that there is no FWM response if $\gamma_{\uparrow\uparrow}$ and $\gamma_{\uparrow\downarrow}$ are negligible. 
In the case $\nu = 1$, the two-exciton states $|\text{X}^{\uparrow\downarrow} \text{X}^{\uparrow\downarrow}\rangle$ do not contribute. 
We discuss the impact of EID on the polarimetric and temporal characteristics of this model in the following. 
Solving the optical Bloch equations for the six-level system in third-order perturbation with respect to the excitation light fields (Supplement S4), 
we arrive at the following temporal dependence of the PE signal measured in the configuration HRH 
    \begin{align}
        \begin{aligned}
    E_\text{HRH} &\propto \text{e}^{-2\Gamma_2 \tau_{12}} \left\{\cos^2(\varphi)\left[1 - (1 - \nu)^2 \text{e}^{-\gamma_{\uparrow\uparrow}\tau_{12}}\right] \right. \\
    & + \frac{1}{2} \left.\left[(1 - \nu)^2 \text{e}^{-\gamma_{\uparrow\uparrow}\tau_{12}} - \text{e}^{-\gamma_{\uparrow\downarrow}\tau_{12}}\right]\right\},\label{eq: two_X_HRH} 
        \end{aligned}
\end{align}
consisting of a term proportional to $\cos^2\varphi$ and a polarization-independent part unless $\nu = 0$ and $\gamma_{\uparrow\uparrow} = \gamma_{\uparrow\downarrow}$. 
Two special cases are the configurations HHH and HVH 
\begin{subequations}
\begin{align}
E_\text{HHH} &\propto \text{e}^{-2\Gamma_2 \tau_{12}} \left\{1 - \frac{1}{2}\left[(1 - \nu)^2\text{e}^{-\gamma_{\uparrow\uparrow}\tau_{12}} + \text{e}^{-\gamma_{\uparrow\downarrow}\tau_{12}}\right]\right\}\\
     E_\text{HVH} &\propto \text{e}^{-2\Gamma_2 \tau_{12}} \frac{1}{2} \left\{(1 - \nu)^2\text{e}^{-\gamma_{\uparrow\uparrow}\tau_{12}} - \text{e}^{-\gamma_{\uparrow\downarrow}\tau_{12}}\right\}.\label{eq: two_X_HVH}
\end{align}
\end{subequations}
Here, the signal in HHH configuration has a component arising from the one-exciton to ground state transitions, which decays at the unperturbed decoherence rate $\Gamma_2$. Instead, the signal in HVH solely stems from the two-exciton to one-exciton transitions and decays faster as given by the rates $\gamma_{\uparrow\uparrow}$ and $\gamma_{\uparrow\downarrow}$. 
Note however, due to the different signs of the contributions resulting from the anti-parallel and parallel spin configurations, a finite rise time or, depending on the relative size of $\gamma_{\uparrow\uparrow}$ and $\gamma_{\uparrow\downarrow}$, a sign inversion of the HVH signal at a non-zero delay $\tau_{12}$ is predicted by the model. 
As follows from the qualitative shape of the polar rosette presented in Figure~\sref{fig: fig02}{d}, the signals in HHH and HVH share the same sign since no zero crossing 
for an intermediate polarization angle $0 \leq \varphi \leq \pi / 2$ is observed. 
This behavior is compatible with the model if the term $\propto \text{e}^{-\gamma_{\uparrow\uparrow}\tau_{12}}$ dominates over the term 
$\propto \text{e}^{-\gamma_{\uparrow\downarrow}\tau_{12}}$ in Equation~\eqref{eq: two_X_HVH}. 
This condition is realized in two scenarios: 
First, correlated states with two excitons of opposite spins $|\text{X}^\uparrow \text{X}^\downarrow\rangle$ are absent in the narrow bandwidth of \qty{0.4}{\milli\eV} of our excitation pulses. 
Second, the state $|\text{X}^\uparrow \text{X}^\downarrow\rangle$ decays significantly faster as the ones with equal spins $|\text{X}^\uparrow \text{X}^\uparrow\rangle$ / $|\text{X}^\downarrow \text{X}^\downarrow\rangle$, i.e. $\gamma_{\uparrow\downarrow} \gg \gamma_{\uparrow\uparrow}$, which is in agreement with the picture of spin-dependent EID. 
In both scenarios, Equation~\eqref{eq: two_X_HRH} can be simplified to 
    \begin{align}
    \begin{aligned}
    E_\text{HRH} &\propto \text{e}^{-2\Gamma_2 \tau_{12}} \left\{\cos^2(\varphi)\left[1 - (1 - \nu)^2\text{e}^{-\gamma_{\uparrow\uparrow}\tau_{12}}\right] \right. \\
     & \left. + \frac{(1 - \nu)^2}{2}\text{e}^{-\gamma_{\uparrow\uparrow}\tau_{12}}\right\}.
    \label{eq: two_X_simplifed_HRH}
    \end{aligned}    
\end{align}
Here, the signal in HVH ($\varphi = \pi / 2$) decays with a rate of $\Gamma_2 + \gamma_{\uparrow\uparrow} / 2$ whereas the main contribution of the signal in HHH ($\varphi = 0$) decays at the unperturbed rate $\Gamma_2$. 
A finite rise time of the signal is not expected in both polarization configurations. 
In this way, the ratio HVH/HHH experiences decaying dynamics given by  
    \begin{equation}
        \text{HVH / HHH} = \left[\frac{2\text{e}^{\gamma_{\uparrow\uparrow} \tau_{12}}}{(1 - \nu)^2}  - 1 \right]^{-1}.   
        \label{eq: HHH_HVH_ratio_2X}
    \end{equation}
Here, it is seen that the ratio between the signal in HVH and HHH for $\tau_{12} = 0$ is determined by the parameter
$\nu$ whereas its decay is defined by $\gamma_{\uparrow\uparrow}$. Thus, even in the case of 
weak EID ($\gamma_{\uparrow\uparrow} \ll \Gamma_2$), comparable magnitudes of the signals in HHH and HVH can be expected.
The model Eq.~\eqref{eq: HHH_HVH_ratio_2X} is fitted to the corresponding experimental data for a fluence of \qty{0.3}{\mu\joule\centi\meter^{-2}} which yields $\gamma_{\uparrow\uparrow} = \qty{5.6(5)}{\mu\eV}$ and $\nu = \num{0.30(1)}$. 
As shown in Figure~\sref{fig: eid_2X_model}{b}, the resulting curve can excellently describe the experimental data for the ratio HVH/HHH. 
Furthermore, the polar rosettes in configurations HRH and HRV for $\tau_\text{ref} = \qty{20}{\pico\second}$ are well described by 
Equations~\eqref{eq: two_X_simplifed_HRH} and (S19b) with the obtained values for~$\gamma_{\uparrow\uparrow}$ and $\nu$, Figure~\sref{fig: eid_2X_model}{c}.
The model thus is successful in describing both the polarization-dependent decoherence rates at a given laser fluence 
as well as the observed qualitative shape of the polar rosettes. 
The derived value of the parameter~$\nu$, approximately \num{0.3}, establishes a reasonable scenario wherein excitons are localized, 
thereby exhibiting a substantial phase-space filling effect ($\nu \neq 0$), yet they are not entirely isolated ($\nu \neq 1)$. 
This is in contrast to well-isolated excitons in (In,Ga)As quantum dots.
The estimated magnitude of the constant $\gamma_{\uparrow\uparrow}$ ranges from $\qty{5.6(5)}{\mu\eV}$ at an exciton density  
of \qty{0.1e17}{\centi\meter^{-3}} up to $\qty{15(1)}{\mu\eV}$ at the highest measured density of \qty{1.4e17}{\centi\meter^{-3}}. 
Thus, the relative impact of exciton-exciton interactions on the homogeneous linewidth of localized excitons is significant.
As a consequence, the optical nonlinearity associated with the spin-dependent interaction in the system under study is of comparable magnitude as the phase-space filling mechanism, which manifests itself in a signal of comparable magnitude in the polarization configurations HHH and HVH.

The agreement between the two-exciton model and the experimental observations is found under the assumption that 
interactions among excitons with opposite spins contribute less to the measured exciton linewidth 
on the $\mu$eV energy scale. 
A microscopic explanation for this behavior might be the relaxation to the biexciton state. 
The biexciton state is energetically located below the considered unbound two-exciton state. 
Relaxation of two excitons into this state, for example through phonon emission, may represent the additional 
scattering channel that could explain the spin-dependent exciton interactions presented in this section.   
The influence of unbound zero spin states on the measured long-lived photon echo response is therefore negligible.

\section{Conclusions}
In conclusion, we demonstrated the important role of the spin degree of freedom for the exciton-exciton interactions in the mixed perovskite crystal \sample. 
As shown in the photon echo experiments, the homogeneous linewidth and the strength of excitation-induced dephasing have a pronounced dependence on 
the relative polarization of the excitation pulses. 
The results are explained by an effective two-exciton model where spin-dependent exciton-exciton correlations are taken into
account. The model suggests a picture in which linewidth broadening on the $\mu$eV energy scale is mediated 
through the interaction among excitons with the same spin.  
We evaluated the magnitude of EID in the range of \qty{10}{\mu\eV} for exciton densities up to \qty{1.4e17}{\centi\meter^{-3}}. 
Measurements of such small line broadenings are achieved at the temperature of \qty{2}{\kelvin} where the exciton linewidth is very narrow due to localization at band gap fluctuations and weak phonon scattering. In this regime, the contribution of the spin-dependent broadening is comparable to the one induced by the phase-space filling. 
This opens up the possibility of using polarization-sensitive excitation for spin-selective control of interacting excitons and exciton complexes, e.g. exciton molecules, in bulk perovskite semiconductors.
\FloatBarrier

\section{Associated content}
\textbf{Supporting information} \\ 
Detailed description of the modeling procedure. 

\section{Author information}
\textbf{Corresponding Authors} \\ 
Stefan Grisard: stefan.grisard@tu-dortmund.de \\
Ilya~A.~Akimov: ilja.akimov@tu-dortmund.de\\

\textbf{ORCID}\\
Stefan~Grisard: 0000-0002-5011-3455\\
Artur~V.~Trifonov: 0000-0002-3830-6035\\
Thilo~Hahn: 0000-0002-8099-4556 \\
Tilmann~Kuhn: 0000-0001-7449-9287 \\
Oleh Hordiichuk: 0000-0001-7679-4423\\
Maksym~V.~Kovalenko: 0000-0002-6396-8938\\
Dmitri~R.~Yakovlev: 0000-0001-7349-2745\\
Manfred~Bayer: 0000-0002-0893-5949\\
Ilya~A.~Akimov: 0000-0002-2035-2324

\section{Acknowledgements}
We thank D.~Smirnov and D.~Wigger for useful discussions.
We acknowledge financial support by the Deutsche Forschungsgemeinschaft via the SPP2196 Priority Program (Project No. 506623857)
as well as through the Collaborative
Research Center TRR 142/3 (Grant No. 231447078, Project No. A02).
The work at ETH Zurich was financially supported by the Swiss National Science 
Foundation (grant agreement 186406, funded in conjunction with SPP219 through the DFG-SNSF bilateral program) 
and by ETH Zurich through ETH+ Project SynMatLab.

\newpage

\def\thesection{S\arabic{section}}
\def\theequation{S\arabic{equation}}
\def\thefigure{S\arabic{figure}}
\def\thetable{S\arabic{table}}
\setcounter{equation}{0}
\setcounter{figure}{0}

\section*{Supplementary Information}

In the following, we derive expressions for the polarization-sensitive four-wave mixing optical response at the exciton resonance by employing optical Bloch equations for three distinct scenarios.
In the initial phase (Section S1), we consider the spin degrees of freedom, represented by two optically active and degenerate exciton states. These states are optically addressed by opposite circular polarizations of light, corresponding to exciton spin "up" and "down". Further discussion on other exciton complexes, such as charged excitons (trions), donor-bound excitons, and bound biexcitons, is provided in Section S2.
In the second stage, we introduce spin-dependent nonlinear contributions in the form of local fields. These contributions induce additional dephasing and energy shifts proportional to the exciton populations in specific spin configurations (Section S3). However, this approach fails to explain the polarization-sensitive decoherence times.
In the last stage (Section S4), we employ a two-exciton model where exciton-exciton interactions are accounted for by considering two-exciton states with higher decoherence rates compared to single exciton states. Notably, the decoherence rates of two-exciton states vary significantly between parallel and antiparallel spin configurations. 
This allows us to explain the different coherence times in co- and cross-polarized configurations. 

\subsection{S1 Non-interacting exciton model}

The four-wave mixing response of 
non-interacting excitons can be calculated using the coupled equations of motion for the 
5 independent elements of the $3\times 3$ density matrix $\rho$ of the V-scheme, that are
the microscopic polarizations 
$p^{\pm} = \langle \text{G}|\rho|\text{X}^{\pm}\rangle$, 
the occupations $n^{\pm} = \langle \text{X}^{\pm}| \rho |\text{X}^{\pm}\rangle$ of the 
excited states, as well as the 
spin polarization $s^+ = \langle \text{X}^{+}| \rho |\text{X}^{-}\rangle = (s^-)^*$
\begin{subequations}
\begin{align}
    \frac{d}{dt} p^{\pm} &= i\Delta p^{\pm} - \frac{p^{\pm}}{T_2} - i \Omega^\pm (1 - n^{\mp} - 2n^{\pm}) + i \Omega^\mp s^{\mp} \\ 
    \frac{d}{dt} n^{\pm} &= i \left(p^{\pm} \Omega^{\pm *} - p^{\pm *} \Omega^\pm\right) - \Gamma_1 n^{\pm} \\
    \frac{d}{dt} s^+ &= i (p^{-} \Omega^{+*} - p^{+ *} \Omega^-).   
\end{align}   
\end{subequations}
Here, $\Gamma_2$ is the decoherence rate, $\Gamma_1$ is the population decay rate, 
$\Delta$ denotes the detuning between central laser frequency and exciton transition energy, 
and $\Omega^\pm = \mu E^\pm(t) / \hbar$ are Rabi frequencies with respect to $\sigma^\pm$ polarized components 
of the slow varying electric field amplitude $E^\pm(t)$. $\mu$ denotes the transition dipole moment which we assume to be equal 
for both excitonic transitions. 

To calculate the four-wave mixing response resulting from the third order polarization $p^{(3)}\propto E_1^* E_2^2$, 
we expand the density matrix elements in perturbative order with respect to the electric field. Further, we assume that the system is 
initially fully in the ground state, which is 
justified at low temperatures of \qty{2}{\kelvin}. In this case, we have to solve the following set of equations to obtain solutions 
for the third-order polarizations   
\begin{subequations}
\begin{align}
    \frac{d}{dt} p^{\pm (1)} &= i\Delta p^{\pm (1)} - \frac{p^{\pm (1)}}{T_2} - i \Omega^\pm \label{eq: p_bloch}\\
    \frac{d}{dt} n^{\pm (2)} &= i \left(p^{\pm (1)} \Omega^{\pm *} - p^{\pm * (1)} \Omega^\pm\right) - \Gamma_1 n^{\pm (2)}\\
    \frac{d}{dt} s^{+ (2)} &= i (p^{- (1)} \Omega^{+*} - p^{+ (1) *} \Omega^-) \\ 
    \frac{d}{dt} p^{\pm (3)} &= i\Delta p^{\pm (3)} - \Gamma_2 p^{\pm (3)} + i \Omega^\pm n^{\mp (2)} + 2i\Omega^\pm n^{\pm (2)} + i \Omega^\mp s^{\mp (2)}.\label{eq: p_3}
\end{align}  
\label{eq: V_scheme_bloch_equations}  
\end{subequations}
We consider the first and second pulses as delta pulses centered at $t = 0$ and $t = \tau_{12}$
\begin{equation}
    \Omega^\pm(t) = \Omega_1^\pm \delta (0) + \Omega_2^\pm \delta (t - \tau_{12}).
\end{equation}
This leads us to the following solutions that fulfill the phase matching condition $2k_2 - k_1$
\begin{equation}
    p^{\pm (3)}(t) = i \Theta(\tau_{12}) \Theta(t - \tau_{12}) \text{e}^{i\Delta (t - 2\tau_{12}) - \Gamma_2 t} \left[(\Omega^\pm_2)^2 \Omega_1^{\pm*} + \Omega_2^\pm \Omega_2^\mp \Omega_1^{\mp*} \right], 
\end{equation}
where $\Theta$ is the Heavi-side function. The macroscopic response of an inhomogeneous ensemble $P^{\pm (3)}\propto \int g(\Delta) p^{\pm (3)}(t, \Delta)$, where $g(\Delta)$ 
is usually a Gaussian distribution of detunings, leading to the photon echo centered at $2\tau_{12}$. 
$P^{\pm (3)}$ represent sources of circularly polarized signal field components $E_S^\pm$.
The polarization characteristics 
of the photon echo amplitude in the configurations HRH and HRV are calculated by constructing linear polarized fields in circular 
polarization base as $E^\pm_i = E_i \exp{\pm i \varphi_i}$, with linear polarization angle $\varphi_i$ and 
total signal field strength $E_i$.
The first pulse is linear polarized, i.e. $\varphi_1 = 0$. The polarization angle $\varphi_2 = \varphi$ of the second pulse is a free parameter. 
We arrive at
\begin{subequations}
\begin{align}
    |E_\text{HRH}| &\propto |E_\text{S}^+ + E_\text{S}^-| \propto \frac{|\mu|^4}{\hbar^3}  E^*_{1} E_{2}^2 \cos^2(\varphi)\\
    |E_\text{HRV}| &\propto |E_\text{S}^+ - E_\text{S}^-| \propto \frac{1}{2} \frac{|\mu|^4}{\hbar^3}  E^*_{1} E_{2}^2 |\sin(2 \varphi)|,
\end{align}
\end{subequations}
i.e., the dependences given in Eq.~(1) of the main text.

\subsection{S2 Discussion of multi-excitonic transitions}\label{sec: multiexciton}

\begin{figure}
    \centering
    \includegraphics[scale = 1]{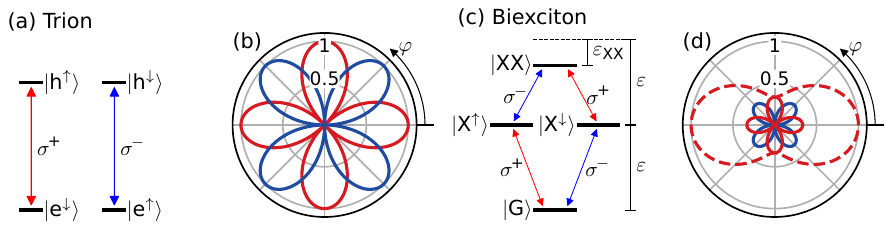}
    \caption{(a)/(c) Energy level arrangements for the trion and biexciton systems. 
    (b)/(d) Polar dependences in the configuration HRH (red) and HRV (blue) for the trion and biexciton 
    systems. For the biexciton, the polar dependence in HRH has a time-dependence 
    as a consequence of the quantum beats at the frequency given by the biexciton binding energy. 
    Solid line: $\tau_{12} = n \hbar / \varepsilon_\text{XX}$; dashed line: $\tau_{12} = (2n + 1) \hbar / 2\varepsilon_\text{XX}$, where 
    $n$ is an integer.}
    \label{fig: T_XX}
\end{figure}

The formation of multi-excitonic complexes potentially has strong implications for the 
polarization selection rules in FWM 
spectroscopy~\cite{yaffe_polarization_1993, haase_coherent_1998, bristow_polarization_2009, suzuki_coherent_2016, poltavtsev_polarimetry_2019, trifonov_photon_2022, grisard_multiple_2022}. 
For example, the binding of excitons to donors or localized carriers can lead to the formation of trions. 
The trion selection rules result from a 4-level scheme shown in Fig.~\sref{fig: T_XX}{a}. Here, both ground 
and excited states are 
two-fold degenerate with electron spin up/down in the ground state and trion spin up/down in the excited state. 
In contrast to the exciton, the two circularly 
polarized transitions are independent of each other. Therefore, the total signal amplitude is fully independent 
of the polarization angle $\varphi$ and the polarization angle of the signal is given by $\varphi_S = 2\varphi$. The 
signal dependences in HRH and HRV are thus simply given by projection on the horizontal and vertical axis $E_\text{HRH} \propto \cos(2\varphi)$ 
and $E_\text{HRV} \propto \sin(2\varphi)$, as visualized in Fig.~\sref{fig: T_XX}{b}. The formation of trions is therefore 
not compatible with our observations. 
The same holds for the bound state of two 
excitons with opposite spin, the biexciton $|\text{XX}\rangle$, 
with a diamond scheme as depicted in Figure~\sref{fig: T_XX}{c}. 
The behavior of the exciton-biexciton system 
within the photon echo polarimetry technique was experimentally 
studied in a \ce{MAPbI3} single crystal in 
Ref.~\cite{trifonov_photon_2022}, where excitons are only weakly localized. 
For the biexciton, the qualitative shape of the HRH dependences has 
a dependence on $\tau_{12}$ as a consequence of the quantum beats at the frequency given by the biexciton binding 
energy~$\varepsilon_\text{XX}$. The two extreme cases for $\tau_{12} = n\hbar/\varepsilon_\text{XX}$ and 
$\tau_{12} = (2n+1)\hbar/2\varepsilon_\text{XX}$, where $n$ is an integer, 
are vizualized in Fig.~\sref{fig: T_XX}{d}. For 
$\tau_{12} = (2n+1)\hbar/2\varepsilon_\text{XX}$, indeed a polar dependence which qualitatively 
resembles our observation is expected, compare dashed line in Fig.~\sref{fig: T_XX}{d} with the experimental 
data in Fig.2(d) in the main text. 
However, the expected quantum beating of the polar dependences is not observed in our experiment, which rules out the 
formation of biexcitons on the time range of several tens of picoseconds. 
We note that the biexciton binding energy in \ce{MAPbI3} is \qty{2.4}{\milli\eV}~\cite{trifonov_photon_2022}, which we 
regard as a lower bound for the biexciton binding energy 
in \sample\, since localization typically leads to increased biexciton binding energy~\cite{cundiff_semiconductor_2008}. 
As our laser pulses are spectrally narrow, \qty{0.4}{\milli\eV}, the formation of biexcitons is not probable. 
However, further investigations using spectrally broad pulses are necessary to experimentally determine the biexciton 
binding energy. 

\subsection{S3 Spin-dependent excitation-induced nonlinearities} \label{sec: eid}
We modify the optical Bloch equations for the exciton V-scheme~\eqref{eq: V_scheme_bloch_equations} to account for spin-dependent EID and EIS. 
Therefore, we introduce excitation-induced decay rates and resonance frequencies as 
\begin{subequations}
    \begin{align}
        \Gamma_2^{\pm} &= \Gamma_2 + \alpha_{\uparrow\uparrow} n^{\pm} + \alpha_{\uparrow\downarrow} n^{\mp} \\
        \omega^{\pm} &= \omega_0 + \beta_{\uparrow\uparrow} n^{\pm} + \beta_{\uparrow\downarrow} n^{\mp}
    \end{align}
\end{subequations}        
where $\Gamma_2$ denotes 
the intrinsic decoherence rate and $\alpha_{\uparrow\uparrow}$/$\alpha_{\uparrow\downarrow}$ (real and positive) are interaction constants 
for excitons with the same and opposite spin. Analogously, $\beta_{\uparrow\uparrow}$/$\beta_{\uparrow\downarrow}$ account for 
excitation-induced shifts for excitons with the same or opposite spin. 
Note that, instead of considering the exciton linewidth to be proportional to the 
total exciton density within the inhomogeneously broadened ensemble, 
we introduce EID separately for each resonance energy. 
This treatment is a prerequisite for the formation of a photon echo signal and 
is well-justified for localized systems~\cite{borri_nonlinear_1997}. 
We introduce complex interaction constants $V = \alpha + i\beta$ to compactly write the modified 
equation of motion for the polarizations (Eq.~\eqref{eq: p_bloch})
\begin{equation}
    \frac{d}{dt} p^{\pm} = i\Delta p^{\pm} - \Gamma_2 p^{\pm} - i \Omega^\pm (1 - n^{\mp} - 2n^{\pm}) + i \Omega^\mp s^{\mp} - (V_{\uparrow\downarrow} n^{\mp} + V_{\uparrow\uparrow} n^\pm )p^\pm \label{eq: p_eid}.
\end{equation}    
Considering the third-order polarization, we modify Eq.~\eqref{eq: p_3} to 
\begin{equation}
    \frac{d}{dt} p^{\pm (3)} = i\Delta p^{\pm (3)} - \Gamma_2 p^{\pm (3)} + i \Omega^\pm n^{\mp (2)} + 2i\Omega^\pm n^{\pm (2)} + i \Omega^\mp s^{\mp (2)} - (V_{\uparrow\downarrow} n^{\mp (2)} + V_{\uparrow\uparrow} n^{\pm (2)} )p^{\pm (1)}.
\end{equation}
Here, it can be seen that through the excitation-dependent decoherence rate and/or frequency shift,
another source term for a third order polarization $\propto  n^{(2)}p^{(1)}$ arises. Considering the phase-matching condition of our experiment, 
such third-order polarization arises from the scattering of the linear polarization created by the second pulse scattered on the 
population grating created by the first and 
second pulses. The total solution for the two third-order exciton polarizations can be written as 
\begin{align}
    \begin{aligned}
    p^{\pm (3)}(t) = \text{e}^{i\Delta (t - 2\tau_{12}) - \Gamma_2 t} &\left[(\Omega^\pm_2)^2 \Omega_1^{\pm*}\left(1 + \frac{V_{\uparrow\uparrow}}{2\Gamma_1}\left(1 - \text{e}^{-\Gamma_1 (t - \tau_{12})}\right) \right)\right. \\
    &\left. + \Omega_2^\pm \Omega_2^\mp \Omega_1^{\mp*}\left(1 + \frac{V_{\uparrow\downarrow}}{2\Gamma_1}\left(1 - \text{e}^{-\Gamma_1 (t - \tau_{12})}\right) \right) \right],
    \end{aligned}
\end{align}   
which leads to the following dependences of the electric field strength in the configurations HRH and HRV
\begin{subequations}
    \label{eq: rosettes_EID}
\begin{align}
    |E_\text{HRH}^\text{EID}| &\propto \frac{|\mu|^4}{\hbar^3}  E^*_{1} E_{2}^2 \text{e}^{-2\Gamma_2\tau_{12}}\left\{\cos^2(\varphi)\left(1 + \frac{V_{\uparrow\uparrow}}{2\Gamma_1}\left(1 - \text{e}^{-\Gamma_1 \tau_{12}} \right)\right) + \frac{V_{\uparrow\downarrow} - V_{\uparrow\uparrow}}{4\Gamma_1}\left(1 - \text{e}^{-\Gamma_1 \tau_{12}} \right)\right\}\\
    |E_\text{HRV}^\text{EID}| &\propto  \frac{|\mu|^4}{\hbar^3}  E^*_{1} E_{2}^2 \text{e}^{-2\Gamma_2\tau_{12}} \frac{|\sin(2\varphi)|}{2}\left\{1 + \frac{V_{\uparrow\uparrow}}{2\Gamma_1}\left(1 - \text{e}^{-\Gamma_1 \tau_{12}} \right)\right\}.
\end{align}
\end{subequations}
Here, we can distinguish between spin-independent and spin-dependent excitation-induced effects. 
For spin-independent EID and EIS, i.e. $V_{\uparrow\uparrow} = V_{\uparrow\downarrow}$, the signal amplitudes are modified while the qualitative shape of the polar dependences is unchanged with respect to the model of non-interacting excitons. 
For spin-dependent EID and EIS, i.e. $V_{\uparrow\uparrow} \neq V_{\uparrow\downarrow}$, a horizontally polarized signal, solely arising from spin-dependent interaction is expected for linearly cross-polarized excitation $\propto V_{\uparrow\downarrow} - V_{\uparrow\uparrow}$ independent of $\varphi$ which fits our observations presented above. 
We thus can conclude that indeed spin-dependent exciton-exciton interactions can qualitatively explain the modified polarimetric behavior of the exciton resonance. We discuss in the following if the simplified expansion of the Bloch equations can also quantitatively reproduce the polarimetric and temporal behavior that we observe.

The excitation-induced contributions predicted by Equations~\eqref{eq: rosettes_EID} scale with the ratio between the interaction constants $V$ and the population decay rate $\Gamma_1$. 
Signals of comparable magnitude in the configuration HHH and HVH thus require a strong influence of EID/EIS. This prediction is a contradiction to the observation presented in Fig.~1(e) of the main text where EID represents only a small correction to the intrinsic linewidth.
Furthermore, all excitation-induced contributions in Equations~\eqref{eq: rosettes_EID} share a prefactor $\propto (1 - \exp(-\Gamma_1 \tau_{12}))$ that rises on a timescale of $1 / \Gamma_1 = T_1 \approx 100$\,ps~\cite{grisard_long-lived_2023}.
In the experiment, no deviations from single exponential decays on that timescale are observed, compare photon-echo decays in Fig.~1(d) of the main text. 
Note that we neglected higher-order terms with respect to the external optical field which lead to a power-dependent decay of the excitation-induced signal. 
Even when high-order terms are taken into account, the described model is not capable of explaining the polarization-sensitive decoherence times that we observed in 
Fig.1(d) of the main text.
We thus can conclude that the simplified consideration of interaction effects between excitons through population-dependent linewidths and shifts is capable of accounting for the additional channels of signal formation but results in wrong statements of the observed 
temporal dynamics as well as the relative amplitudes of signals in different polarization configurations.

\subsection{S4 Four-wave-mixing response of two-exciton model} \label{sec: XX_model}
In this section, we describe the calculation of the four-wave-mixing response of the two-exciton model depicted in 
Fig.~3(a) of the main text. We follow a standard perturbative multi-wave mixing expansion of the density matrix of the 
system as described for example in Refs.~\citenum{mukamel_1999, smallwood_analytical_2017}. 

We represent the six states of the two-exciton model as $|\text{G}\rangle\hat{=}$ (1, 0, 0, 0, 0, 0)$^\top$, $|\text{X}^\uparrow\rangle\hat{=}$ (0, 1, 0, 0, 0, 0)$^\top$, 
$|\text{X}^\downarrow\rangle\hat{=}$ (0, 0, 1, 0, 0, 0)$^\top$, $|\text{X}^\uparrow \text{X}^\downarrow\rangle\hat{=} $(0, 0, 0, 1, 0, 0)$^\top$, 
$|\text{X}^\uparrow \text{X}^\uparrow\rangle\hat{=}$ (0, 0, 0, 0, 1, 0)$^\top$, and $|\text{X}^\downarrow \text{X}^\downarrow\rangle\hat{=}$ (0, 0, 0, 0, 0, 1)$^\top$.  
The dynamics of the corresponding $6\times 6$ density matrix ${\rho}$ is determined by the Liouville-von Neumann Equation 
\begin{equation}
    \frac{d}{dt}{\rho} = \frac{i}{\hbar} \left[{\rho}, \mathbf{H} + \mathbf{V}(t) \right].
    \label{eq: liouville}
\end{equation} 
Here, $\textbf{H}$ is the Hamiltonian of the bare system (written in rotating frame)
\begin{equation}
    \textbf{H} =  \text{diag}(0, \hbar\Delta, \hbar\Delta, 2\hbar\Delta, 2\hbar\Delta, 2\hbar\Delta)
\end{equation}
with the detuning $\Delta = \omega_L - \omega_0$ between the laser frequency $\omega_L$ and the transition frequency $\omega_0$. 
The interaction with light is described by the matrix $\mathbf{V}(t)$
\begin{footnotesize}
\begin{equation}
    \mathbf{V}(t) = \begin{pmatrix}
        0 & \mu E^*_+(t) & \mu E^*_-(t) & 0 & 0 & 0 \\ 
        \mu E_+(t) & 0 & 0 & \mu E^*_-(t) & \sqrt{2}(1 - \nu)\mu  E^*_+(t) & 0 \\ 
        \mu E_-(t) & 0 & 0 & \mu E^*_+(t) & 0 & \sqrt{2}(1 - \nu)\mu  E^*_-(t) \\ 
        0 & \mu E_-(t) & \mu E_+(t) & 0 & 0 & 0 \\
        0 & \sqrt{2}(1 - \nu)\mu  E_+(t) & 0 & 0 & 0 & 0 \\
        0 & 0 & \sqrt{2}(1 - \nu)\mu  E_-(t) & 0 & 0 & 0 \\
    \end{pmatrix}
    \label{eq: V_matrix}
\end{equation}
\end{footnotesize}
where 
\begin{equation}
    E_\pm(t) = E_{\pm, 1} \text{e}^{i \mathbf{k}_1 \cdot \mathbf{r}} \delta(0) + E_{\pm, 2} \text{e}^{i \mathbf{k}_2  \cdot \mathbf{r}} \delta(\tau_{12}).
\end{equation}
with the circular polarized electric field components $E_{\pm, \text{n}}$ of the n-th pulse with wave vector $\mathbf{k}_\text{n}$. The temporal envelopes of the 
first and second pulses are 
assumed as delta functions $\delta(t)$ centered at 0 and $\tau_{12} > 0$, respectively.       
The dipole moments of the transitions in Eq.~\eqref{eq: V_matrix} are motivated in the main text.  

We assume that the two optical pulses are weak such that we can expand the density matrix according to Equation~\eqref{eq: liouville} into perturbative orders 
${\rho}^{(\text{m})}$.   
The m-th order is given by
\begin{equation}
    \rho^{(\text{m})}_\text{ij}(t) = 
  \int_{-\infty}^t \frac{i}{\hbar} \left [{\rho}^{(m -1)}(\tau), \mathbf{V}(\tau) \right ]_\text{ij} \text{e}^{-i \Omega_\text{ij} (t - \tau)} d\tau,
  \label{eq: pn_expansion}
  \end{equation}
where $\Omega_\text{ij} = (H_{\text{ii}} - H_{\text{jj}}) / \hbar - i \Gamma_\text{ij}$ 
with
\begin{equation}
    {\Gamma} = \begin{pmatrix}
        0 & \Gamma_2 & \Gamma_2 & 0 & 0 & 0 \\ 
        \Gamma_2 & 0 & 0 & \Gamma_2 + \gamma_{\uparrow\downarrow} & \Gamma_2 + \gamma_{\uparrow\uparrow} & 0 \\ 
        \Gamma_2 & 0 & 0 & \Gamma_2 + \gamma_{\uparrow\downarrow} & 0 & \Gamma_2 + \gamma_{\uparrow\uparrow} \\ 
        0 & \Gamma_2 + \gamma_{\uparrow\downarrow} & \Gamma_2 + \gamma_{\uparrow\downarrow} & 0 & 0 & 0 \\
        0 &\Gamma_2 + \gamma_{\uparrow\uparrow} & 0 & 0 & 0 & 0 \\
        0 & 0 & \Gamma_2 + \gamma_{\uparrow\uparrow} & 0 & 0 & 0 \\
    \end{pmatrix},
    \label{eq: Gamma_matrix}
\end{equation}
accounting for phenomenological decoherence rates as introduced in the main text. Note that population decay rates are neglected since they do not contribute 
to the two-pulse photon echo signal. We assume that the system is initially fully in the ground state, i.e. $\rho^{(0)}_{11}(0) = 1$, which is justified 
considering the low temperature of \qty{2}{\kelvin} of our experiments.

Here, we only consider the lowest perturbative order resulting in a photon echo response $\text{m} = 3$ (four-wave mixing).
From the full solution of ${\rho}^{(3)}$ given by Eq.~\eqref{eq: pn_expansion}, we select those terms that fulfill the phase matching condition of 
our experiment $2\mathbf{k}_2 - \mathbf{k}_1$. 
An arbitrary polarization component of the four-wave mixing signal is constructed from the circular polarization components $E^\pm$ 
\begin{align}
    E^+ &\propto \mu \left(\rho^{(3)}_{12} + \rho^{(3)}_{34}\right) + \sqrt{2} (1 - \nu) \mu \rho^{(3)}_{25} \\
    E^- &\propto \mu \left(\rho^{(3)}_{13} + \rho^{(3)}_{24}\right) + \sqrt{2} (1 - \nu) \mu \rho^{(3)}_{36}. 
\end{align}
In the main text, we discuss the polarization configurations HRH and HRV, where the first pulse is linearly polarized, the second pulse's polarization is rotated 
by an angle $\varphi$, and the signal is detected in horizontal/vertical polarization. The circular polarization components of the second pulse are thus constructed as 
$E_{\pm, 2} \propto \text{exp} (\pm i\varphi)$. 
We arrive at the following expressions for the photon echo amplitude as a function of delay $\tau_{12}$ and polarization angle~$\varphi$
\begin{subequations}
\begin{align}
        E_\text{HRH} &\propto \text{e}^{-2\Gamma_2 \tau_{12}} \left\{\cos^2(\varphi)\left[1 - (1 - \nu)^2 \text{e}^{-\gamma_{\uparrow\uparrow}\tau_{12}}\right] + \frac{1}{2} \left[(1 - \nu)^2 \text{e}^{-\gamma_{\uparrow\uparrow}\tau_{12}} - \text{e}^{-\gamma_{\uparrow\downarrow}\tau_{12}}\right]\right\}  \label{eq: two_X_HRHz} \\
        E_\text{HRV} &\propto \text{e}^{-2\Gamma_2 \tau_{12}} \frac{\sin(2\varphi)}{2} \left\{1 - (1 - \nu)^2\text{e}^{-\gamma_{\uparrow\uparrow}\tau_{12}} \right\},\label{eq: two_X_HRV}
\end{align}
\end{subequations}
where Eq.~\eqref{eq: two_X_HRHz} is given as Eq.~(4) in the main text.

\providecommand{\latin}[1]{#1}
\makeatletter
\providecommand{\doi}
  {\begingroup\let\do\@makeother\dospecials
  \catcode`\{=1 \catcode`\}=2 \doi@aux}
\providecommand{\doi@aux}[1]{\endgroup\texttt{#1}}
\makeatother
\providecommand*\mcitethebibliography{\thebibliography}
\csname @ifundefined\endcsname{endmcitethebibliography}
  {\let\endmcitethebibliography\endthebibliography}{}

\end{document}